\documentclass[12pt,english]{article}
\pdfoutput=1
\usepackage{babel}
\usepackage{mathtools,amssymb,cite,slashed}
\usepackage[width=16cm,height=23cm,centering]{geometry}
\usepackage{setspace}
\usepackage[hidelinks,pagebackref]{hyperref}
\usepackage{cleveref,ytableau,tikz}

\numberwithin{equation}{section}
\newcommand\beq{\begin{equation}}
\newcommand\eeq{\end{equation}}
\newcommand{\nnn}[1]{\tikz[remember picture] \node[] (#1) {};}

\newcommand {\BS} {\mathbb S}
\newcommand {\BC} {\mathbb C}
\newcommand {\BR} {\mathbb R}
\newcommand {\BZ} {\mathbb Z}
\newcommand {\ii} {\mathrm i}
\newcommand {\by} {\mathbf y}
\newcommand {\bz} {\mathbf z}
\newcommand {\ba} {\mathbf a}
\newcommand {\bm} {\mathbf m}
\newcommand {\CalM} {\mathcal M}
\newcommand {\CalL} {\mathcal L}
\newcommand {\CalZ} {\mathcal Z}
\newcommand {\CalH} {\mathcal H}
\newcommand {\ve}  {\varepsilon}
\DeclareMathOperator{\Cone} {Cone}
\DeclareMathOperator{\Res} {Res}
\DeclareMathOperator{\Hom} {Hom}
\DeclareMathOperator{\End} {End}
\DeclareMathOperator{\tr} {tr}
\DeclareMathOperator{\rk} {rk}
\renewcommand{\hat}{\widehat}
\renewcommand{\Re} {\operatorname{Re}}

\title{Magnificent Four with Colors}
\author{Nikita Nekrasov\footnote{On leave from Kharkevich IITP (Moscow) and Alikhanov ITEP (Moscow)} \and Nicol\`o Piazzalunga}
\date{ \it\small Simons Center for Geometry and Physics\\Stony Brook University, Stony Brook NY 11794-3636 USA}
\begin{document}
\maketitle

\abstract{We study the rank $N$ magnificent four theory, which is the supersymmetric localization of $U(N)$ super-Yang-Mills theory with matter (a super-group $U(N|N)$ gauge theory) on a Calabi-Yau fourfold. Our theory contains the higher rank Donaldson-Thomas theory of threefolds. We conjecture an explicit formula for the  partition function $\CalZ$, and report on the  performed checks.
The partition function $\CalZ$ has a free field representation. Surprisingly, it depends on the Coulomb and mass parameters in a simple way.  We also clarify the definition of the instanton measure.}

\tableofcontents

\section{Introduction and summary}

In this paper we define the higher rank version of the `magnificent four' theory\cite{Nekrasov:2017cih}, and conjecture the free field expression
of its partition function.  

The latter is essentially the twisted Witten index of the supersymmetric quantum mechanics describing a collection of $k$ $D0$-branes in the presence of  
$N$ $D8$-branes and anti-branes wrapping a Calabi-Yau four-fold,
which we take to be ${\BC}^4$,
in the background of the appropriate $B$-field\cite{Witten:2000mf}.  Of course, the collection of both branes and anti-branes in flat space in the absence of additional fields breaks all supersymmetry. The open string spectrum contains a tachyon that drives the system towards a supersymmetric ground state, in which the branes and anti-branes annihilate.
When $B$-field is present,
the complete annihilation need not take place. In our problem we turn on additional background fields. 

The intrigue of these problems is that they might teach us about the degrees of freedom of $M$-theory. Recall that the analogous setup
involving $D0$- and $D6$-branes \cite{Nekrasov:2004vv, Nekrasov:2005bb, Nekrasov:2014nea} reproduces the twisted
index of eleven-dimensional supergravity viewed as quantum field theory (when expanded around the Minkowski space) with global supersymmetry.\footnote
{Up to a square puzzle, yet to be resolved, not to be confused with the square puzzle of Refs.~\cite{Ooguri:2004zv, Beasley:2006us}.}
In our present case the introduction of higher dimensional branes might probe additional degrees of freedom. 

We shall assume the low-energy description of our system of branes and anti-branes is given by an $U(N|N)$ gauge theory, as in Ref.~\cite{Vafa:2001qf}, although in our case this is a cohomological field theory, not the Chern-Simons one.

As we shall see, mathematically our problem provides an interesting (albeit not the uniform) measure on the space of colored solid partitions.
Even for rank 1, the counting of such objects is not known,
namely the generating function of solid partitions
\begin{multline}
P_{3}(q) := \sum _{k=0}^{\infty } p_{3}(k)\ q^{k}= 1+q+4\ q^{2}+10\ q^{3}+26\ q^{4}+59\ q^{5}+140\ q^{6}+\cdots \\
+ 214071 q^{16} + \cdots 
\end{multline}
where $p_3(k)$ counts the number of solid partitions of size $k$,
does not have a known closed form;
only its values up to $k \sim 72$
and its asymptotic behavior for large $k$\cite{Destainville:2014nra} are known. See Ref.~\cite{Knuth} for the computational aspects of $P_{3}(q)$.

From the gauge-theoretic viewpoint, the $D0-D8$ system provides a generalization of the ADHM construction
for some $8+1$ dimensional gauge theory living on a stack of $D8$-branes (with the anti-$D8$'s playing the role of defects providing some sort of fundamental matter fields). 
Namely, our $D0$-branes probing $N$ $D8$-branes provide the generalized $U(N)$ instantons, whose energy is concentrated in codimension $8$, in agreement with Ref.~\cite{Douglas:1995bn}.

The relevant Chern-Simons coupling on the $D8$ worldvolume $X^{9}$ is
\beq
\frac{1}{2(2{\pi}{\ii})^{2}} \, \int_{X^{9}} C^{(5)} \wedge \tr F \wedge F
\eeq 
where $F$ is the  field strength  of the $U(N)$ gauge field living on the stack of $N$ $D8$-branes and $C^{(5)}$ the RR 5-form. Now suppose
$X^{9} = {\BR}^{1} \times Y^{8}$, with a compact Calabi-Yau fourfold $Y^{8}$, and take $C^{(5)} = dt \wedge \Re \Omega$
with $\Omega$ a holomorphic volume form on $Y^{8}$. 
In this way one gets the action of the semi-topological gauge theory of  Ref.~\cite{Baulieu:1997jx}.

Finally, by restricting to the case of rank $1$ and by taking the cohomological limit, which corresponds to high temperature limit $\beta \to 0$ in the quantum mechanical problem,
one makes contact with Donaldson-Thomas invariants of Calabi-Yau four-folds recently discussed in Ref.~\cite{Cao:2017swr}.

\paragraph{Organization of the paper}

\Cref{sec:d8} deals with the $D0-D8$-brane system: in \cref{sec:lag} we present the susy quantum mechanics,
which is used to compute the index in \cref{sec:comp},
after recalling some relevant definitions related to 4d partitions in \cref{sec:part};
\cref{sec:tang} reformulates the problem from the viewpoint of representation theory of tangent space,
in particular in \cref{sec:mov} we prove that the tangent space is movable, and in \cref{sign-rule} we prove a sign rule that allows one to pass from the residue picture to the tangent space picture;
in \cref{sec:conj} we conjecture an expression for the instanton partition function.
We conclude with a discussion of CS-terms in \cref{sec:cs} and some aspects of wall-crossing phenomena in \cref{sec:wc}.
In \cref{sec:d6} we comment on how our conjecture encompasses many previously studied cases and how its various limits reproduce previously known results.
Finally, in \cref{5inst} we provide a simple example of computation,
in \cref{example} we check our statement in the simplest non-trivial case, 
and in \cref{defs} we recall some useful definitions.
\Cref{sec:ale} contains some useful facts about orbifold characters.

\paragraph{Acknowledgements}
NN thanks A.~Okounkov for discussions, and MSRI (Berkeley), Department of Mathematics (UC Berkeley), IHES (Bures-sur-Yvette), and Skoltech (Moscow) for hospitality during the preparation of this work and for opportunity to present its preliminary results there.
Research of NN is partially supported by Laboratory of
Mirror Symmetry NRU HSE, RF Government grant, ag.~N.~14.641.31.0001.
NP thanks A.~Grassi, A.~Waldron and the participants of the workshops at ICTP Trieste (July 2018) and GGI Florence (April 2018), where this work was presented, for stimulating discussions.
The research of NP was supported in part by Istituto Nazionale di Alta Matematica ``Francesco Severi'' (INdAM)
and by Perimeter Institute for Theoretical Physics.\footnote
{Research at Perimeter Institute is supported by the Government of Canada through the Department of Innovation, Science and Economic Development and by the Province of Ontario through
the Ministry of Research and Innovation.}
This is a post-peer-review, pre-copyedit version of an article published in {\it Communications in Mathematical Physics}.
The final authenticated version is available \href{https://dx.doi.org/10.1007/s00220-019-03426-3}{online}.

\section{The \texorpdfstring{$D0-D8$}{D0-D8} brane partition functions}
\label{sec:d8}

In this section we formulate the matrix quantum mechanical problem aimed at representing the system of $D0$-branes in the background of $D8$- and anti-$D8$-branes. Even if the brane interpretation of our 
matrix quantum mechanics is questionable, its twisted Witten index is a well-defined quantity that we compute below. 

\subsection{Matrix quantum mechanics}

The system of $k$  $D0$-branes slowly moving in flat Minkowski space is described by the maximally supersymmetric gauged quantum mechanics, which is obtained by the dimensional reduction of ten dimensional super-Yang-Mills theory to $0+1$ dimensions. 

The wavefunctions of this quantum system are the $U(k)$-invariant functions of  $9$ bosonic  and $8$ fermionic 
Hermitian $k \times k$ (adjoint-valued) matrices. Indeed, the sixteen-component Majorana-Weyl fermions become, upon quantization, the generators
of real Clifford algebra, half of which  could be chosen as odd coordinates (with the additional parity constraints).  The $U(1)$-part describes the helicity components
of the Kaluza-Klein modes of the supergraviton in eleven dimensions compactified onto ${\BS}^{1} \times {\BR}^{1,9}$, while the non-abelian
$SU(k)$ part is believed to have a unique $L^2$-normalizable supersymmetric ground state \cite{Witten:1995im}. 

The presence of additional $N$ $Dp$-branes adds to the matter content the $k \times N$ and $N \times k$-valued matrix fields, bosonic and fermionic, depending on the brane's spatial dimensionality $p$ and the choice of GSO projection.

The quantity of our primary interest is the twisted Witten index
\beq
{\CalZ}_{N,k}^{Dp} (g) = {\tr}_{{\CalH}_{k,N}} \, (-1)^{F} g\, e^{-\beta {\hat H}_{k,N}}
\label{eq:twwi}
\eeq 
where $g$ is an element of the global symmetry group, which commutes with at least one of the supercharges. 
Of course, \cref{eq:twwi} depends only on the conjugacy class of $g$.

\subsubsection{Lagrangian and supersymmetry}
\label{sec:lag}

The low energy action for $k$ $D0$-branes is the dimensional reduction of super-Yang-Mills theory from $9+1$ to $0+1$ dimensions.
We are interested in $D0$-branes moving in the background of $D8$-branes with $B$-field, therefore we organize the fields according to the symmetry breaking $Spin(9,1) \to Spin(8) \times Spin(1,1)$.
The global symmetry commuting with the two preserved supercharges is $SU(4) \subset Spin(8)$.
Our supersymmetric quantum mechanics has gauge group $G=U(k)$. The bosonic field content is given by the $4$ complex $k\times k$ matrices $B^{a}$, their conjugates $B^{\dagger}_{a}$, $a = 1, 2, 3, 4$, 
a real vector $A_t$ and a real scalar $\Phi$, all valued in the adjoint representation of $G = U(k)$. We denote by $K$ the defining $k$-dimensional representation of $G$. 
We rotate into Euclidean signature (as \cref{eq:twwi} corresponds to the periodic Euclidean time) and combine them into $\sigma= A_t + {\ii}\Phi$, $\sigma ^\dagger = A_t - {\ii}\Phi$.
The covariant derivative $\partial_t + {\ii} A_t$ can be complexified to $D_t = \partial_t + {\ii} \sigma$.
The $Spin(9,1)$ Mayorana-Weyl spinor decomposes, under $Spin(8)$,  as ${\bf 8}_s+{\bf 8}_c$; in turn, 
these decompose as ${\bf 8}_s = {\bf 4} + {\bar{\bf 4}}$ and ${\bf 8}_c = {\bf 6} +{\bf 1}+{\bf 1}$ under $SU(4) \subset Spin(8)$.

{}Accordingly, the fermions are $\psi^{a}$, $\psi_a^\dagger$, $\chi^{ab} = - {\chi}^{ba} =  \frac 12 {\varepsilon}^{abcd} \chi_{dc}^\dagger$, $\eta$ and $\chi$ all valued in $\Pi \End(K)$;
the auxiliary bosons $h^{ab}= - h^{ba} =  \frac 12 \varepsilon^{abcd} h_{dc}^\dagger$ and $h=h^\dagger$ are valued in $\End(K)$.

The $N$ $D8$-branes along ${\BR}_t\times {\BR}^8$ contribute
$D0-D8$ strings giving bosons $I \in \Hom (N,K)$ and $I^\dagger \in \Hom(K,N)$ from the Neveu-Schwarz sector,
with fermionic partners $\psi \in \Pi \End(N,K)$ and $\psi^\dagger \in \Pi \End(K,N)$ from the Ramond sector.
Here $N$ is the $N$-dimensional vector space corresponding to Chan-Paton indices.
The fields $I$ and $I^\dagger$ are present only when we turn on the $B$-field,
which we take to be $B = \sum_{a=1}^4 b_a \, dx_{2a-1}\wedge dx_{2a}$.

We also introduce an auxiliary system of
fermions $\Upsilon \in \Pi \Hom (M,K)$, $\Upsilon^\dagger \in \Hom(K,M)$
and bosons $H \in \Hom(M,K)$.
This \emph{may} come from R sector of $N$ anti-$D8$-branes,
whose Chan-Paton bundle we denote by $M$.

Let us combine the two preserved supercharges into a supercharge $\delta$, which squares to $\delta^2 = D_t$.
The susy algebra reads
\beq
\begin{aligned}
\delta B^a = \psi^a, 
\quad \delta I = \psi \, , \quad 
\delta B^{\dagger}_{a} = \psi^{\dagger}_a\, ,
\quad \delta I^{\dagger} = \psi^{\dagger} \, , \\
\delta  \psi^a = D_{t}B^{a} = {\partial}_{t}B^{a} + {\ii} [{\sigma}, B^{a}],
\quad \delta  \psi = D_{t} I = {\partial}_{t} I + {\ii} {\sigma} I \, , \\
\delta  \psi^{\dagger}_{a} = D_{t}B^{\dagger}_{a} = {\partial}_{t}B^{\dagger}_{a} - {\ii} [{\sigma}, B^{\dagger}_{a}],
\quad \delta  \psi^{\dagger} = D_{t} I^{\dagger} = {\partial}_{t} I^{\dagger} - {\ii}  I^{\dagger} {\sigma}  \\
\quad \delta \chi^{ab} = h^{ab} , \quad {\delta}h^{ab} = D_{t}{\chi}^{ab} = {\partial}_{t}{\chi}^{ab} + {\ii} [ {\sigma}, {\chi}^{ab} ] \, , \\
\quad \delta \chi = h \, , \quad {\delta}h = D_{t}{\chi} = {\partial}_{t}{\chi} + {\ii} [ {\sigma}, {\chi} ] \\
 \delta \sigma^\dagger = \eta \, ,  \quad {\delta}{\eta} = D_{t}{\sigma}^{\dagger} = {\partial}_{t}{\sigma}^{\dagger} + {\ii} [{\sigma}, {\sigma}^{\dagger} ] \, ,  \\
\delta \Upsilon = H\, , \quad {\delta}H = D_{t}{\Upsilon} = {\partial}_{t}{\Upsilon} + {\ii} {\sigma} {\Upsilon} \, , 
 \\
\delta \sigma = 0 \, .
\end{aligned}
\eeq 
Define
\beq
s^{ab} = [B^a,B^b]+ \frac12 \varepsilon^{abcd} [B_d^{\dagger}, B_c^{\dagger}] = \frac12 \varepsilon^{abcd} s_{cd}^\dagger, \qquad
\mu = \sum_{a=1}^4 [B^a,B_a^\dagger] + I I^\dagger - 1_k \zeta
\eeq 
The action is $S = \int dt \, {\CalL}$, ${\CalL} = {\delta}  {\bf\Psi}$, where
\begin{multline}
{\bf\Psi} =  {\tr}\left( \psi^\dagger {\bar D}_{t} I + {\psi} {\bar D}_{t} I^{\dagger} \right) + \sum_a {\tr} \left( \psi_a^\dagger {\bar D}_{t} B^a + {\psi}^{a} {\bar D}_{t} B^{\dagger}_{a} \right) + {\tr} \left( \eta D_{t}{\Phi} \right)    + \\
   + \sum_{a<b} {\tr} \chi_{ab}^\dagger \left( \frac 12 h^{ab} - s^{ab} \right) + {\tr} \chi \left( \frac 12 h - {\mu} \right) 
 + \frac 12 {\tr} \left( \Upsilon^\dagger H + H^{\dagger}{\Upsilon} \right)
\end{multline}
The FI term $\propto \zeta>0$ represents the effect of the $B$-field.  The classical Higgs branch $\CalM$ is the zero set of the scalar potential
\beq
V = \frac 12 {\tr} {\mu}^2 + \frac 12 \sum_{a<b} {\tr} s_{ab}^\dagger s_{ab}
\eeq 
modulo $U(k)$ gauge transformations.
On the space of solutions to the equations we have $[B^a,B^b]=0$ and $\tr I I^\dagger=\zeta k>0$. One shows \cite{Nekrasov:2017cih} that 
the equation $\mu = 0$ implies the stability condition
\beq
K =  {\BC} \left[ B^1,B^2,B^3,B^4 \right] \, I( N )\ 
\label{eq:stab}
\eeq 
Conversely, the $GL(K)$-orbit of any tuple $(B^{a}, I)$ obeying \cref{eq:stab} crosses the locus 
${\mu}=0$ (which is, in turn, invariant under the $U(k)$-action). 

The real virtual dimension of $\CalM$ is $2Nk$, the number of components of $\Upsilon$.
Our partition function \cref{eq:twwi} can be written as some sort of $\chi_y$-genus as
\beq
\CalZ_{N,k}^{D8} (g) = \int_{[\CalM]^\text{virt}} \hat A \, \operatorname{ch} \wedge^\bullet E
\eeq
where $E=\Hom (K,M)$ is a vector bundle over $\CalM$ and $\operatorname{ch} \wedge^\bullet E = \sum_{i=0}^{\rk E} (-1)^i \operatorname{ch} \wedge^i E$.
Both the A-roof and the Chern character are defined equivariantly, and contain the respective equivariant weights.

\subsection{Localization}

Our system has the manifest global symmetry group $G_F = U(N)_{\rm color} \times SU(4) \times U(N)_{\rm flavor}$,
which acts on the bundles $N$, $K$ and $M$.
Denoting the fields collectively by $\phi$,
the Witten index \cref{eq:twwi} is computed by the path integral
\beq
\int_\text{TBD} [d \phi] \, e^{-\int_{0}^{\beta} {\CalL} \, dt}
\eeq 
where TBD denotes the space of fields obeying the twisted boundary conditions $\phi(\beta) = g \phi(0)$ for $g \in G_F$.
The $U(1)$-valued coordinates on the Cartan torus of $G_F$ are the exponentiated
Coulomb branch parameters $\{ {\nu}_\alpha\}_{\alpha=1}^{N} \in U(1)^{N} \subset U(N)_{\rm color}$, the exponentiated
fundamental masses $\{ \mu_\alpha\}_{\alpha=1}^{N} \in U(1)^{N} \subset U(N)_{\rm flavor}$
and the rotational $\Omega$-background parameters $(q_1,\ldots,q_4) \in U(1)^{3} \subset SU(4) \subset Spin(8)$,
where $q_a$ rotates the two dimensional plane $\langle x_{2a-1},x_{2a} \rangle$ inside $\BR^8$, with the $SU(4)$ constraint
\beq
Q:= \prod_{a=1}^4 q_a =1
\eeq
enforced to preserve the two supercharges. 

In what follows we will denote the vector spaces with the $G_F$-action as their characters, e.g.~$M=\sum \mu_\alpha$ and $N=\sum {\nu}_\alpha$.

Instead of dealing with the twisted boundary conditions, one can 
weakly gauge $G_F$, i.e.\ give a vev to the complexified\footnote
{One needs to uplift the quantum mechanics to the two dimensional sigma model to geometrize the scalar in the vector multiplet of $G_F$.}
background gauge field $A_t + {\ii} {\Phi}$ that can be gauge transformed to make it valued in the complexified Cartan subalgebra of $G_F$.\footnote
{The condition ${\partial}_{t}{\Phi} + {\ii} [A_t, {\Phi}] = 0$ is required to preserve supersymmetry.}
The complexification of the Cartan subgroup $T_{F}$ of $G_{F}$ is parameterized by the  ${\BC}^{\times}$-valued parameters ${\nu}_\alpha = e^{{\ii} a_\alpha}$,
$q_a = e^{{\ii} {\ve}_a}$, $\mu_\alpha = e^{{\ii} m_\alpha}$, so that ${\bz} = \{ {\ba} = (a_{\alpha})_{\alpha = 1}^{N}, {\ve} = ({\ve}_{a})_{a=1}^{4}, {\bm} = (m_{\alpha})_{{\alpha}=1}^{N}  \}\in {\mathfrak t}_F^{\BC}$ spans the complexified Cartan subalgebra of $G_F$.

Denote the new supercharge by $\delta_{\bz}$ and the new Lagrangian by ${\CalL}_{\bz} = \delta_{\bz} \mathbf \Psi_{\bz}$. 
Specifically, 
\beq
\begin{aligned}
\delta_{\bz} B^a = \psi^a, 
\quad \delta_{\bz} I = \psi \, , \quad 
\delta_{\bz} B^{\dagger}_{a} = \psi^{\dagger}_a\, ,
\quad \delta_{\bz} I^{\dagger} = \psi^{\dagger} \, , \\
\delta_{\bz}  \psi^a = D_{t}B^{a} + {\ii} {\ve}_{a} B^{a}\, , 
\quad \delta_{\bz}  \psi = D_{t} I - {\ii} I {\ba} \, , \\
\delta_{\bz}  \psi^{\dagger}_{a} = D_{t}B^{\dagger}_{a} - {\ii} {\ve}_{a} B^{\dagger}_{a} \, , 
\quad \delta_{\bz}  \psi^{\dagger} = D_{t} I^{\dagger} + {\ii}{\ba} I^{\dagger} \\
\quad \delta_{\bz} \chi^{ab} = h^{ab} , \quad \delta_{\bz} h^{ab} = D_{t}{\chi}^{ab} + {\ii} ({\ve}_{a}+{\ve}_{b}) {\chi}^{ab} \, , \\
\quad \delta_{\bz} \chi = h \, , \quad \delta_{\bz} h = D_{t}{\chi} \, , \quad
 \delta_{\bz} \sigma^\dagger = \eta \, ,  \quad \delta_{\bz} {\eta} = D_{t}{\sigma}^{\dagger}  ,  \\
\delta_{\bz} \Upsilon = H\, , \quad \delta_{\bz} H = D_{t}{\Upsilon} - {\ii} {\Upsilon}{\bm} \, , \ \delta_{\bz} \Upsilon^{\dagger} = H^{\dagger}\, , \quad \delta_{\bz} H^{\dagger} = D_{t}{\Upsilon}^{\dagger} + {\ii}{\bm} {\Upsilon} \, , 
 \\
\delta_{\bz} \sigma = 0 \, , 
\end{aligned}
\label{eq:defdel}
\eeq 
and
\begin{multline}
{\bf\Psi}_{\bz} =  {\tr}\left( \psi^\dagger \left( {\bar D}_{t} I + {\ii} I {\bar\ba} \right) + {\psi} \left( {\bar D}_{t} I^{\dagger} - {\ii} {\bar\ba} I^{\dagger}\right)  \right) +\\
+  \sum_a {\tr} \left( \psi_a^\dagger \left( {\bar D}_{t} B^a   - {\ii}{\bar\ve}_{a} B^{a} \right) + {\psi}^{a} \left( {\bar D}_{t} B^{\dagger}_{a} + {\ii} {\bar\ve}_{a} B^{\dagger}_{a} \right) \right) + {\tr} \left( \eta D_{t}{\Phi} \right)    + \\
   + \sum_{a<b} {\tr} \chi_{ab}^\dagger \left( \frac 12 h^{ab} - s^{ab} \right) + {\tr} \chi \left( \frac 12 h - {\mu} \right) 
 + \frac 12 {\tr} \left( \Upsilon^\dagger H + H^{\dagger}{\Upsilon} \right)
\end{multline}
where  in the unitary theory ${\bar\ba}$ and ${\bar\ve}_{a}$ are complex conjugates of
$\ba$ and $\ve_{a}$, respectively.
The index \cref{eq:twwi} is now computed by the path integral with periodic boundary conditions\cite{Cecotti:1981fu, Sethi:1997pa, Moore:1998et, Smilga:2001my}
\beq
\label{eq:twwi2}
\CalZ_{N,k}^{D8} (g) = \int_\text{PB} [d\phi] \, e^{- \int_{0}^{\beta} {\CalL}_{\bz} \, dt} 
\eeq

\subsubsection{Solid partitions}
\label{sec:part}

We can think of higher-dimensional partitions recursively. Start from a Young diagram:
this is a collection $Y=(\ell_1,\ldots,\ell_s)$ with $s\geq 1$ of positive integers $\ell_i$
such that $\ell_i \geq \ell_{i+1}$ for $i=1,\ldots,s-1$,
and we denote its size by $|Y|=\sum_{i=1}^s \ell_i$.
Inclusion is defined as $Y \subseteq Y'$ iff $\ell_i \leq \ell_i'$ for all $i$.
The next step is a plane partition: this is collection $\pi=(Y_1,\ldots,Y_s)$
of Young diagrams $Y_i$
such that $Y_{i+1} \subseteq Y_i$.
Inclusion is defined as $\pi \subseteq \pi'$ iff $Y_i \subseteq Y_i'$ for all $i$'s,
and the size is $|\pi|=\sum_{k=1}^s |Y_k|$.
The next step is a solid partition: this is a collection $\rho=(\pi_1,\ldots,\pi_s)$ of plane partitions $\pi_i$,
such that $\pi_{i+1} \subseteq \pi_i$.
Its size is $|\rho|=\sum_{k=1}^s |\pi_k|$.
Equivalently we can think of a solid partition $\rho$ as a collection of non-negative integers $\{ \rho_{i,j,k} \}$ indexed by integers $i,j,k \geq 1$ subject to the conditions
$\rho_{i,j,k} \geq \rho_{i+1,j,k}$, $\rho_{i,j,k} \geq \rho_{i,j+1,k}$ and $\rho_{i,j,k} \geq \rho_{i,j,k+1}$. The size is $|\rho| = \sum_{i,j,k} \rho_{i,j,k}$.
In this formulation, we can regard the solid partition $\rho$ as the subset of points $(a,b,c,d) \in \BZ^4$, such that $a,b,c,d \geq 1$ and $d \leq \rho_{a,b,c}$.
The character of a solid partition $\rho$ is
\beq
\operatorname{ch}_\rho (q_1, q_2, q_3, q_4)= \sum_{(a,b,c,d) \in \rho} q_1^{a-1} q_2^{b-1} q_3^{c-1} q_4^{d-1}
\eeq

A colored solid partition $\vec \rho = (\rho_1,\ldots,\rho_N)$ is a $N$-dimensional vector of solid partitions, where we call $N$ the rank.
We define its character as
\beq
K = \sum_{{\alpha}=1}^N {\nu}_{\alpha} \operatorname{ch}_{\rho_{\alpha}} (q_1, q_2, q_3, q_4)
\eeq 
where we introduced Coulomb branch parameters ${\nu}_{\alpha}$. Its size is $|\vec \rho|=\sum\limits_{{\alpha}=1}^N |\rho_{\alpha}|$.

We define the dual $K^*$ of $K$ by replacing $q_a$ with $q_a^{-1}=q_a^*$ for $a \in \{1,2,3,4\}$
and similarly for $N$ and $M$.
We call a finite sum (with $\pm 1$ coefficients) of finite products of $K$, $N$, $M$ and their duals a virtual character.
A virtual character is \emph{movable} when it does not contain $\pm 1$ factors in the sum.

\subsubsection{Ordering of monomials}

We introduce the following version of the lexicographic order on monomials in four variables $(q_1, q_2, q_3, q_4)$:
before imposing the condition $Q=1$, first order monomials by increasing powers of $q_4$, then order terms with same power of $q_4$ by increasing powers of $q_3$,
then order terms with the same power of $q_4$ and the same power of $q_3$ by increasing powers of $q_2$, and finally do the same with $q_1$;
in formulas
\begin{multline}
(i,j,k,l) < (a,b,c,d) \, \Leftrightarrow \\
(l <d ) \vee (l =d \wedge k < c) \vee (l=d \wedge k =c \wedge j < b) \vee (l=d \wedge k=c \wedge j=b \wedge i<a)
\end{multline}
where $\vee$ stands for OR, while $\wedge$ stands for AND.

Once an ordering has been chosen, the content of a solid partition $\rho$ of size $k$ is
\beq
x_\rho = (q_1^ {a-1} q_2^ {b-1} q_3^ {c-1} q_4^ {d-1})_{(a,b,c,d) \in \rho}
\eeq 
where the $A$-th component of $x_\rho$ comes from the $A$-th quadruple $(i,j,k,l)\in \rho$.
We associate to $\rho$ the flag
\beq
F_\rho = [ \{ x_1=x_{\rho,1} =1 \} \supset \{x_1=x_{\rho,1} \} \cap \{ x_2=x_{\rho,2} \} \supset \cdots \supset \{ x=x_\rho \}]
\eeq 
which can be used to define the iterated residue
\beq
\Res_{F_\rho}  =  \Res_{x_k=x_{\rho,k}} \cdots \Res_{x_2=x_{\rho,2}} \Res_{x_1=x_{\rho,1}}
\eeq

\subsection{Computation of the index}
\label{sec:comp}

Let $\mathfrak t$ be the Cartan subalgebra of $G=U(k)$,
$k =\rk G$,
$W$ its Weyl group,
and  $u \in \mathfrak t^\BC$.
The fields $B^a$ and $I$ provide a representation of $G \times G_F$
\beq
V_\text{chiral}
= \bigoplus_{i \in \mathcal I} \BC (\mathcal Q_i, \mathcal Q_i^F)
= \bigoplus_{i,j=1}^k \bigoplus_{a=1}^4  \BC (Q_{ij},Q_a^F) \oplus \bigoplus_{\ell=1}^k \BC(Q_\ell)
\eeq 
where we use calligraphic letters for a generic theory and italic ones for our theory.
Here $\mathcal Q_i \in \mathfrak t^*$ and $\mathcal Q_i^F \in \mathfrak t_F^*$ are the weights of the $i$-th field,
namely $\mathcal Q_i(u), \mathcal Q_i^F(\bz) \in \BC$.
Pick a basis $\{e_1,\ldots,e_k\}$ of $\mathfrak t^*$ where $e_a-e_b$ are roots of $SU(k)\subset U(k)$ and $e_i(u)=u_i$,
and a basis $\{E_1,\ldots,E_4\}$ of $\BR^{4*}$ where $E_a-E_b$ are roots of $SU(4)\subset G_F$ and $E_a(\bz)={\ve}_a$.
Then $Q_{ij}= e_i-e_j$, $Q_i=e_i$ and $Q_a^F=E_a-1/4 \sum_{b=1}^4 E_b$.
For $i \in \mathcal I$ let
\beq
H_i := \{ u \in \mathfrak t^\BC \,| \, \mathcal Q_i(u) + \mathcal Q_i^F(\bz)=0 \}
\eeq 
Denote $\mathcal I_s$ the set labeling (distinct) hyperplanes, $\mathcal Q = \{ \mathcal Q_i | i \in \mathcal I_s \}$ and
\beq
\mathfrak M_\text{sing} := \{ u \in \mathfrak t^\BC \,| \, k \text{ linearly independent hyperplanes meet at } u \}
\eeq 
where two hyperplanes are linearly independent iff their normal vectors are linearly independent
(iff the corresponding $\mathcal Q_i$'s are linearly independent).
For $u_* \in \mathfrak M_\text{sing}$,
let $\mathcal I_s (u_*)$ be the subset of $\mathcal I_s$ indexing hyperplanes that contain $u_*$,
and correspondingly let $\mathcal Q(u_*) = \{ \mathcal Q_i | i \in \mathcal I_s (u_*) \}$.
For the case at hand, we always have
\beq
Q(u_*) \subseteq \{ e_i | \, 1 \leq i \leq k \} \cup \{ e_i-e_j | \, 1 \leq i,j \leq k , i \neq j\}
\eeq

The fields $\chi^{ab}$ and $\Upsilon$ provide another representation of $G\times G_F$
\beq
V_\text{fermi}= \bigoplus_{j \in \mathcal J} \BC(\mathcal R_j,\mathcal R^F_j)
= \bigoplus\limits_{i,j=1}^k \bigoplus\limits_{a < b}^3 \BC( Q_{ij},Q^F_{ab}) \oplus \bigoplus_{i=1}^k \BC(Q_i,Q^{(m)})
\eeq 
with $\mathcal R_i \in \mathfrak t^*$ and $\mathcal R_i^F \in \mathfrak t_F^*$.
We have $Q^F_{ab} = E_a + E_b-\frac12 \sum\limits_{c=1}^4 E_c$ and $Q^{(m)} (\bz) = m$.
If we denote $\alpha$ the roots of $G$ and set $x_i=e^{2\ii u_i}$, $y_i = e^{2\ii z_i}$,
from the one-loop determinants we get the rational function
(set $x^{\mathcal Q_i} = e^{2\ii \mathcal Q_i (u)}$ and $[w]:=w^{1/2}-w^{-1/2}$)
\beq
\begin{aligned}
\chi_k &=
\prod_\alpha [x^\alpha] \frac { \prod_{j \in \mathcal J} [x^{\mathcal R_j} y^{\mathcal R_j^F} ]} {\prod_{i \in \mathcal I} [x^{\mathcal Q_i} y^{\mathcal Q_i^F} ]} \\
& =
\prod_{i >j}^k \frac {\sin^2 (u_i -u_j) \prod_{a<b}^4 \sin (u_i -u_j -{\ve}_a-{\ve}_b)} {\prod_{a=1}^4 \sin (u_i - u_j -{\ve}_a) \sin( u_j - u_i - {\ve}_a) }
\prod\limits_{i=1}^k \prod_{{\alpha}=1}^{N} \frac {\sin (u_i - m_{\alpha})} { \sin (u_i - a_{\alpha})}
\end{aligned}
\eeq 

Under the assumptions that
(i) $\zeta\not \in \Cone_\text{sing} \mathcal Q$,
(ii) for every $u_*$ in $\mathfrak M_\text{sing}$ the set $\mathcal Q (u_*)$ is projective,
and (iii) the space of fields $\{ \phi_i \}_{i \in \mathcal I}$ that have a zero mode at $u=u_*$ and solve the D-term constraints is compact,
the index \cref{eq:twwi2} is\cite{Hori:2014tda}
\beq
\CalZ_{N,k}^{D8} = \frac 1 {|W|} \sum_{u_* \in \mathfrak M_\text{sing}} \operatorname{JK-Res}_{u_*, \zeta} \chi_k \, d^k u
\eeq 
where
the JK-Residue gives a prescription for which poles $u_*$ should be included
that depends on $\zeta$ through the chamber of $\mathfrak t^*$ w.r.t.~$\mathcal Q(u_*)$ to which $\zeta$ belongs.

Dimension of Weyl group is $|W|=k!$ for $U(k)$.
$\chi_k$ is invariant under Weyl group permuting the $u_i$'s and
under $S_4$ acting on the $q_a$'s.\footnote
{If we simultaneously invert all $q_a$'s to $q_a^{-1}$, $\chi_k$ picks a factor of $(-1)^k$.}

If we can find a $\mathcal{FL}(\mathcal Q (u_*)) ^+$-regular $\xi$ in the same chamber of $\mathfrak t^*$ w.r.t.~$\mathcal Q(u_*)$ as $\zeta$,
the JK-Residue is equal\cite{SzenesVergne2,SzenesVergne} to a sum of iterated residues computed at flags $F \in \mathcal{FL}^+ (\mathcal Q(u_*),\xi)$,
which has been studied from the algebro-geometric viewpoint by Parshin\cite{mazin}.
In our case, if we take $\zeta = +(e_1+\cdots +e_k)$ and
\beq
\xi = \zeta + \delta \sum_{i=1}^k (k-i+1) e_{\pi(i)}
\eeq 
for some small $\delta>0$ and some permutation $\pi$,
then only one flag contributes.

Let us take a different route. Fix $u_* \in \mathfrak M_\text{sing}$ and put
\beq
\Sigma := \left\{ \sigma =\{ \sigma_1,\ldots,\sigma_k\} \subseteq Q(u_*) | \, \sigma \text{ spans } \mathfrak t^* \right\}
\eeq 
The JK-residue for $f_\sigma = \frac 1 {\prod_{r=1}^k \sigma_r (u) }$, $\sigma \in \Sigma$ is
\beq
\operatorname{JK-Res} f_\sigma =
\begin{cases} 1, & \zeta \in \Cone (\sigma) \\
0, & \text{otherwise} \end{cases}
\eeq 
Near $\tilde u :=u-u_*=0$ we can write\footnote
{One should use functions $\tilde f_\sigma = \prod_{r=1}^k \sin^{-1} \sigma_r(u)$. However since $\tilde f_\sigma \sim f_\sigma$ near $\tilde u=0$, the distinction is  unnecessary.}
(in a non-unique way and possibly up to terms that do not contribute to the residue)
\beq
\label{expansion}
\chi_k = \sum _ {\sigma \in \Sigma'} f_\sigma h_\sigma
\eeq 
for some $\Sigma' \subseteq \Sigma$ and holomorphic functions $h_\sigma$ that are non-zero near $\tilde u=0$:
the function $\chi_k$ has at least $k$ vanishing factors at denominator (of the form $(x_j-1)$ or $(x_i-x_j)$),
assume $k$ of these are a basis for the ideal of linear functions vanishing at $\tilde u=0$
(otherwise the residue vanishes), so we can use it to eliminate one vanishing factor at numerator;
for each of the remaining summands, we repeat the above argument, and we are left with \cref{expansion};
this is the only possible contribution to the residue.

If $u_*$ contributes to the residue, then there must exist $\sigma \in \Sigma$ such that $\zeta \in \Cone (\sigma)$.
For $\zeta = - (e_1+\ldots +e_k)$ there is no choice of $\sigma\in\Sigma$ satisfying the above condition, and the result is zero.
For $\zeta = + (e_1+\ldots +e_k)$ there are many such sets that meet the condition.
For example for $k=2$ the sets $\{ e_1, e_2\}$, $\{ e_1,e_2-e_1\}$ and their permutations are allowed,
with $f_{ \{ e_1, e_2\} } = \frac 1 {u_1 u_2}$ and $f_{ \{ e_1,e_2-e_1\} } = \frac 1 {u_1 (u_2-u_1)}$.
For generic $k$, an admissible set $\sigma$ must contain at least one $e_j$, which implies $e_i-e_j$ is allowed while $e_j -e_i$ is not;
if it contains $e_j$ and $e_b - e_j$, then it can contain $e_a-e_b$ but not $e_b-e_a$;
therefore each set has a tree structure.

Among the allowed poles $u_*$, some may still not contribute to the residue due to cancellations among numerator and denominator.
We want to prove by induction on $k$ that the existence of $\sigma \in \Sigma'(u_*)$
with $u_* \in \mathfrak M_\text{sing}$ admissible
implies that $u_*$ is a solid partition.
By Weyl symmetry, we can always assume $\exists i < k$ such that
\beq
\sigma = \sigma^{(k-1)} \cup \{ e_k - e_i \}
\eeq
where $\sigma^{(k-1)}$ only contains $e_1,\ldots, e_{k-1}$ and
is admissible with respect to the truncation $u_*^{(k-1)}$.
Since in the decomposition
\beq
\chi_k (x_1, \ldots, x_k ) = \delta \chi_k (x_1,\ldots, x_k) \chi_{k-1} (x_1, \ldots, x_{k-1} )
\eeq
the factor $\sigma^{(k-1)}$ can only come from $\chi_{k-1}$,
we have $\sigma^{(k-1)} \in \Sigma'(u_*^{(k-1)})$, $u_*^{(k-1)}$ admissible.
Therefore by the induction hypothesis $u_*^{(k-1)}$ is a solid partition.

\begin{figure}[h]
\centering
\begin{ytableau}
\nnn{n1} & *(black!25) \nnn{n2} \\
\nnn{n3} & \nnn{n4} & & \\
\nnn{n5} & & &
\end{ytableau}
\begin{tikzpicture}[remember picture, overlay]
\draw[thick] (n2) -- (n1);
\draw[thick] (n2) -- (n4);
\draw[thick] (n3) -- (n1);
\draw[thick] (n3) -- (n4);
\draw[thick] (n3) -- (n5);
\draw[thick,double] (n2) -- (n3);
\end{tikzpicture}
\caption{A partition with some links in $d=2$.}
\label{hypercube}
\end{figure}
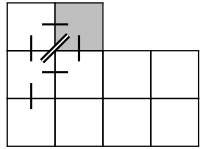

We now show that the allowed ways to go from $k-1$ to $k$ correspond to the possible ways of making a size $k$ solid partition out of a size $k-1$ one.
Let us examine the various possibilities, first without imposing $Q=1$:
\begin{itemize}
\item if $u_k=u_j + {\ve}_c = u_{j'}$ for some $j'<k$, namely if the new box $u_k$ is already present in the partition,
then $u_k -u_j-{\ve}_c=u_k -u_{j'}$ and the singular term at denominator is canceled by a term at numerator
\item if $u_k = u_j + {\ve}_c = u_{j'}+{\ve}_d$ for some $d \neq c$ and $j'<k$, then there exists some $j''$ such that
$u_k = u_{j''} + {\ve}_c + {\ve}_d$ (this argument can be repeated pairwise)
\item if there exists $ a$ such that $u_k -{\ve}_a$ does not belong to the partition, then $u_k -{\ve}_a \neq u_i$ for all $i<k$, and
since $u_j = u_\ell +{\ve}_a$ for some $\ell<k $, then $u_k = u_\ell + {\ve}_a + {\ve}_c$.
\end{itemize}
Such correspondence is not altered when taking the $Q \to 1$ limit,
as we argue by looking at \cref{hypercube} (and prove later in \cref{sign-rule}).
A gray box corresponds to a possible way of adding a box to our partition,
a single line linking two boxes corresponds to a pole (a box in dimension $d$ has at most $2d$ of these),
a double line can either correspond to a zero (when the two boxes differ by ${\ve}_a + {\ve}_b$ for some $a$ and $b$, at most $\binom{d}{2}$ possibilities)
or link boxes that are placed along the same diagonal of some hypercube
and therefore differ by $m \sum_{a=1}^d {\ve}_a$ for some integer $m$
(\cref{hypercube} is in $d=2$ where these two notions coincide, but in $d>2$ they are distinct).
In $d=4$ when we impose $Q=1$, all the boxes located along the same diagonal as the gray box in the picture
start interacting with each other, but the counting is such that no new poles are created.

Finally, at a pole $u_*$ given by a solid partition $\rho$, by linearity of JK-residue and thanks to the fact that
\beq
\operatorname{JK-Res} f_\sigma = \Res_{F_{\rho,\pi}} f_\sigma
\eeq 
for all $\sigma \in \Sigma$,
where the iterated residue is computed at the flag
\beq
F_{\rho,\pi} = [ \{ x_{\pi(1)} = x_{\rho,1} =1\} \supset \{x_{\pi(1)} =x_{\rho,1} \} \cap \{ x_{\pi(2)}=x_{\rho,2} \} \supset \cdots \supset \{ x=x_\rho \}]
\eeq 
determined by the ordered content of the solid partition and some permutation $\pi$ that depends on $u_*$,
we have
\beq
\label{jktoit}
\operatorname{JK-Res} \chi_k = \Res_{F_{\rho,\pi}} \chi_k
\eeq 
Weyl invariance of $\chi_k$ implies RHS of \cref{jktoit} is invariant
if we replace a permutation $\pi_1$ by another one $\pi_2$,
and therefore the $k!$ factor is canceled in the summation over fixed points.

\subsubsection{The colored measure}

For rank $N$ and size $|\vec \rho|=\sum_i |\rho_i| = k$, let $q_{ij}=q_i q_j$ and 
\beq
\# = q_4^{k^2} \left( \frac{ \prod_{1 \leq a<b \leq 3} (1-q_{ab}) }{ \prod_{a=1}^4 (1-q_a) } \prod_{\alpha=1}^N \sqrt{\frac{{\nu}_\alpha}{{\mu}_\alpha}} \right)^k
\eeq 
\beq
\chi_k = \#
\prod_{1 \leq i \neq j \leq k} \frac{(x_j-x_i) \prod_{1 \leq a<b \leq 3} (x_j-x_iq_{ab}) } { \prod_{a=1}^4 (x_j-x_iq_a) }
\prod_{i=1}^k \prod_{\alpha=1}^N \frac{{\mu}_\alpha-x_i}{{\nu}_\alpha-x_i}
\eeq 
Define the measure
\beq
\label{res-meas}
M_{\bz} (\vec \rho) = \operatorname{Res}_{x=x_{\vec\rho}} \chi_k \, \prod_{i=1}^k \frac{d x_i} {x_i} 
\eeq 
on the space of colored solid partitions,
where the iterated residue is computed at
\beq
x_{\vec \rho}= \left( \nu_i q_1^{a_i-1} q_2^{b_i-1} q_3^{c_i-1} q_4^{d_i-1} \right)_{(a_i,b_i,c_i,d_i) \in \rho_i, \quad i=1,\ldots, N}
\eeq 
after imposing the condition $Q=1$ and using a well-defined ordering.\footnote
{In our experiments the result does not change if we impose $Q=1$ after computing the residue.}

From our discussion above it follows that the instanton partition function
\beq
\CalZ_N^{D8} (p, {\bz}) =
\sum_{k=0}^\infty p^k \CalZ_{N,k}^{D8} ({\bz})
\eeq 
takes the form
\beq
\CalZ_{N,k}^{D8} ({\bz}) = \sum_{|\vec \rho|=k} M_{\bz}(\vec \rho)
\eeq 
The instanton fugacity $p$ can be related to the string coupling and the Ramond-Ramond $U(1)$ gauge field via
\beq
p = \exp \oint_{S_1} \frac {ds}{g_s} + \ii C^{(1)}
\eeq 

\subsection{Mathematical properties of virtual tangent space}
\label{sec:tang}

The character of the virtual tangent space to the moduli space at a fixed point, i.e.\ a colored solid partition, is given by
\beq
\label{tangent}
T = (N-M)^* K - P_{123} KK^*
\eeq 
where $P_A=\prod_{a \in A} (1-q_a)$ for $A \subseteq \{1,2,3,4\}$.
$T$ is a square-root of
\beq
T^2 = (N-M) K^* + K (N-M)^* Q - K K^* P_{1234}
\eeq 
We want to apply the map $\hat a$ to $T$ to get the measure \cref{res-meas} obtained from the residue computation:
the map is such that on monomials $r,s$ it acts as $\hat a(r)=1/[r]$,
mapping sums to products $\hat a(r+s) = \hat a(r) \hat a(s)$,
and it converts $T$ to a product of weights in equivariant K theory.

\subsubsection{The movable feast} 
\label{sec:mov}

We want to prove that the tangent space is movable. It is clear that for generic $a_{\alpha}$'s it is enough to show this for $N=1$. We proceed by induction: for the solid partition of size $1$ it is true. Assume $T$ is movable for all solid partitions of the size $\leq k-1$. Any size $k$ solid partition $\rho$ is obtained by adding a box $(A,B,C,D)$ to the size $k-1$ solid partition ${\rho}^{-}$.  We can safely assume the added box to be the highest, i.e.\ for any 
$(i,j,k,l) \in {\rho}^{-}$, $(i,j,k,l) < (A,B,C,D)$. Let $T^{-} = T_{{\rho}^{-}}$. Then
$T \equiv T_{\rho} = T^{-} + {\xi}^{-1} (1 - {\mu}) - P_{123} {\xi} {\chi}^{*} - P_{123} {\xi}^{-1} {\chi}^{-}$, where
\beq
{\xi} = q_{1}^{A-1}q_{2}^{B-1}q_{3}^{C-1}q_{4}^{D-1}\, , \qquad
{\chi}^{-} = \sum_{(i,j,k,l)\in {\rho}^{-}} q_{1}^{i-1}q_{2}^{j-1}q_{3}^{k-1}q_{4}^{l-1}
\eeq 
We can represent the character of the new partition $\rho$ as a sum of two terms:
\beq
{\chi} = {\chi}^{-} + {\xi} = {\chi}_{{\Pi}({\xi})} + {\chi}_{{\rho}\backslash{\Pi}({\xi})}
\eeq 
where
\beq
{\chi}_{{\Pi}({\xi})} = \sum_{(i,j,k,l)\in {\Pi}({\xi})} q_{1}^{i-1}q_{2}^{j-1}q_{3}^{k-1}q_{4}^{l-1}
 \equiv \sum_{i=1}^{A}\sum_{j=1}^{B}\sum_{k=1}^{C}\sum_{l=1}^{D} q_{1}^{i-1}q_{2}^{j-1}q_{3}^{k-1}q_{4}^{l-1} \, 
 \eeq 
(recall that if $(i,j,k,l) \in\rho$ then $(i-1, j,k,l) \in \rho$, $(i,j-1,k,l) \in \rho$ etc.\ as long as $i, j, \ldots $ obey $i >1$, $j >1$ etc.),
\beq
{\chi}_{{\rho} \backslash {\Pi}({\xi})} = \sum_{(i,j,k,l)\in {\rho}\backslash {\Pi}({\xi})} q_{1}^{i-1}q_{2}^{j-1}q_{3}^{k-1}q_{4}^{l-1}\, ,  \eeq 
where $(i,j,k,l)\in {\rho}\backslash {\Pi}({\xi}) \Longrightarrow (i > A) \vee (j>B) \vee (k>C)$ (recall that for all $(i,j,k,l) \in \rho$, $l \leq D$). 

Denote by $\left[ \chi \right]^{(0)}$, for a virtual character $\chi$, the sum of the unmovable terms, i.e.\ the monomials equal to $1$. 
Then:
\beq
\left[ {\xi}^{-1} (1 - {\mu}) \right]^{(0)} = {\delta}_{A=B=C=D}\, , 
\eeq 
(we assume $\mu$ be generic, i.e.\ not equal to any monomial of the form $q_{1}^{\geq 0}q_{2}^{\geq 0}q_{3}^{\geq 0}q_{4}^{\geq 0}$), 
\beq
\left[ P_{123} {\xi}^{-1} {\chi}_{{\rho}\backslash{\Pi}({\xi})} \right]^{(0)} = \left[ P_{123} {\xi} {\chi}_{{\rho}\backslash{\Pi}({\xi})}^{*} \right]^{(0)} = 0 \, , 
\label{eq:movdelta}
\eeq 
which we prove below, and
\begin{multline}
\left[ P_{123} {\xi}^{-1} {\chi}_{{\Pi}({\xi})} \right]^{(0)} = \left[ {\xi}^{-1} (1-q_{1}^{A})(1-q_{2}^{B})(1-q_{3}^{C}) \sum_{l=1}^{D} q_{4}^{l-1} \right]^{(0)} = 
\\
= \left[ - (1-q_{1}^{-A})(1-q_{2}^{-B})(1-q_{3}^{-C}) \sum_{k=1}^{D} q_{123}^{k} \right]^{(0)} = 
{\delta}_{A=B=C \in \{ 1, \ldots , D \}}
\end{multline}
\begin{multline}
\left[ P_{123} {\xi} {\chi}_{{\Pi}({\xi})}^{*} \right]^{(0)} = \left[ -q_{123} {\xi} (1-q_{1}^{-A})(1-q_{2}^{-B})(1-q_{3}^{-C}) \sum_{l=1}^{D} q_{4}^{1-l} \right]^{(0)} = 
\\
= \left[   (1-q_{1}^{A})(1-q_{2}^{B})(1-q_{3}^{C}) \sum_{l=0}^{D-1} q_{123}^{-l} \right]^{(0)} = 
1- {\delta}_{A=B=C \in \{ 1, \ldots , D-1 \} }
\end{multline}
Let us now prove \cref{eq:movdelta}. First:
\beq
\left[ P_{123}  \sum_{(i,j,k,l) \in {\rho}\backslash{\Pi}({\xi}) \wedge k > C} q_{1}^{A-i}q_{2}^{B-j}q_{3}^{C-k} q_{123}^{l-D} \right]^{(0)} = 0 \, , 
\eeq 
since $k > C$ implies $l < D$ (since $l=D$ implies $k \leq C$) so all the monomials in the sum have strictly negative power of $q_3$,
\beq
\left[ P_{123}  \sum_{(i,j,k,l) \in {\rho}\backslash{\Pi}({\xi}) \wedge j > B} q_{1}^{A-i}q_{2}^{B-j}q_{3}^{C-k} q_{123}^{l-D} \right]^{(0)} = 0 \, , 
\eeq 
since $j > B$ implies either $l < D$, in which case all the monomials in the sum have strictly negative powers of $q_2$, or $l =D$, which implies $k < C$, in which case all the monomials
in the sum have strictly positive powers of $q_3$, 
 \beq
\left[ P_{123}  \sum_{(i,j,k,l) \in {\rho}\backslash{\Pi}({\xi}) \wedge i  > A} q_{1}^{A-i}q_{2}^{B-j}q_{3}^{C-k} q_{123}^{l-D} \right]^{(0)} = 0 \, , 
\eeq 
since $i > A$ implies either $l < D$, in which case all such monomials in the sum have strictly negative powers of $q_1$, or $l=D$, which either implies $k < C$, in which case all such monomials
in the sum have strictly positive powers of $q_3$, or $k =C$ and $j < B$, in which case all such monomials in the sum have strictly positive powers of $q_2$. 

Secondly, 
\beq
\left[ P_{123}  \sum_{(i,j,k,l) \in {\rho}\backslash{\Pi}({\xi})} q_{1}^{i-A}q_{2}^{j-B}q_{3}^{k-C} q_{123}^{D-l} \right]^{(0)} = 0 \, , 
\eeq 
since the terms with $k > C$ have strictly positive powers of $q_{3}$, the terms with $j > B$ have strictly positive powers of $q_2$, and the terms with $i > A$ have strictly positive powers of $q_1$.

\subsubsection{The sign rule}
\label{sign-rule}

For simplicity we deal with rank one case.
In this case the single Coulomb branch parameter can be scaled away,
so $K$ is just the character of a solid partition $\rho$ of size $k$, and the only other parameter left from $M$ we call $\mu$:
\cref{tangent} becomes
\beq
T = (1- \mu)^* K - P_{123} K^* K
\eeq 
Construct the formal character $T^f$ by taking $T$ and replacing monomials in $K$ by formal variables $\{x_j\}$, namely
\beq
K \to \sum_{j=1}^k x_j,
\qquad
K^* \to \sum_{j=1}^k x_j^{-1}
\eeq 
By definition of $\hat a$
\beq
M_\bz (\rho) = \Res_{x=x_\rho} \prod_{j=1}^k \frac{dx_j}{x_j} \hat a (T^f +k)
\eeq 
where the residue is well-defined since $T$ is movable and the $(+k)$ term takes care precisely of the $k$ $(-1)$'s that come from $\sum_{j=1}^k x_j x_j^{-1}$.
It is crucial that the identification $x=x_\rho$ follows a well-defined ordering:
this means that we order the monomials in $K$ and assign to each of them a distinct $x_j$ according to such ordering.

We want to prove by induction on $k=|\rho|$ that
\beq
\label{induction}
\Res_{x=x_\rho} \prod_{j=1}^k \frac{d x_j}{x_j} \hat a (T^f +k) =
(-1)^{h(\rho)} \hat a (T)
\eeq 
where
\beq
h(\rho) = |\rho| + \# \{ (a,d) \, | \, (a,a,a,d) \in \rho \text{ and } a \leq d \}
\eeq 
One can check the case $k=1$,
where we fix orientation by requiring $\Res_{x=1} \hat a (x) = +1$.
Suppose \cref{induction} is true for every $\rho$ of size less or equal to $k-1$, and look at a partition of size $k$:
its character will be the sum of a character of a partition of size $k-1$ (call it $\chi$) and a new term (call it $\xi$).
Without loss of generality we can choose $\xi$ such that it comes last in our ordering defined above.
This means that the residue over $x_k=\xi$ is the last one to be computed.

Since $T$ is movable, so is $\delta T := T_{\chi+\xi}-T_\chi$,
where the subscript of $T$ denotes the character of the partition where it is computed,
and by the induction hypothesis
\beq\begin{aligned}
\label{res-split}
& \Res_{\{x_i\}_{i=1}^{k-1}=\chi, \, x_k=\xi} \prod_{j=1}^k \frac{dx_j}{x_j} \hat a (T_\chi^f +\delta T^f +k) = \\
&= (-1)^{h(\chi)} \hat a(T_\chi)
\hat a (1-P_{123})
\Res_{x=\xi} \frac{dx}{x} \hat a \left( (1-\mu^*)x -P_{123}\chi^* x -P_{123}\chi x^{-1} \right)
\end{aligned}
\eeq 
This is so because, thanks to the properties of Parshin residue, by the time we come to computing the residue over $x_k=\xi$,
all the other variables have already been substituted for their value,
and $\delta T^f$ takes the form written in \cref{res-split}.

Let us focus on this last term:
we need to identify possible $\pm 1$'s coming from $x^{-1}$ term as $x \to \xi$:
these will produce the minus signs we are after when combining with possible $\pm 1$'s from $x$ terms,
as we know the argument of residue has exactly one pole as $x \to \xi$
(because $\delta T$ is movable),
and the expression is otherwise identical to $\delta T$.
This task is equivalent to counting the net number of $\pm 1$'s in
$A_1:=-P_{123} \chi \xi^*$.
Parametrize 
\beq
\chi = \sum_{a=1}^h q_4^{a-1} \chi_a(q_1,q_2,q_3)
\eeq 
and distinguish two cases:
\begin{itemize}
	\item case (a): $\xi = q_4^h$. In this case $A_1=-P_{123}\sum_{a=1}^h q_{123}^{h+1-a} \chi_a$ has no $\pm 1$'s.

	\item case (b): $\xi=q_4 ^{h-1} \eta$ with the monomial $\eta \in \chi_{h-1}$ but $\eta \not\in \chi_h$. Of course it must be $\eta/q_i \in \chi_h$ for some $i \in \{1,2,3\}$.
		Let us parametrize $\eta = q_1^{A-1} q_2^{B-1} q_3^{C-1}$, with $A,B,C \geq 1$ and $ABC \geq 2$.
		By considering the parallelepiped $P(\eta)$ generated by $\eta$, we decompose each $\chi_a = \chi^{P(\eta)} + \Delta_a^\eta$,
	where
	\beq
	\chi^{P(\eta)} = \sum_{i=1}^A q_1^{i-1} \sum_{j=1}^B q_2^{j-1} \sum_{k=1}^C q_3^{k-1} 
	\eeq 
	and $\Delta_a^\eta$ is the sum of all terms in $\chi_a$ where at least one power of $q_1$, $q_2$ or $q_3$ is (strictly) bigger than $A-1$, $B-1$ or $C-1$ respectively.
	Notice that since $\eta \not \in \chi_h$, we have to add and subtract it to write $\chi_h+\eta = \chi^{P(\eta)} + \Delta_h^\eta$.
	This decomposition allows to sum over $\chi^{P(\eta)}$, while it is easy to show that $\Delta_a^\eta$ terms do not contribute unmovable terms:
	we can write (up to movable terms)
	\beq
	\begin{aligned}
			A_1 &=-P_{123} \eta^{-1} \sum_{a=1}^h q_{123}^{h-a} \chi_a \\
			&=  -P_{123} \eta^{-1} \sum_{a=1}^{h-1} q_{123}^{h-a} \chi^{P(\eta)} -P_{123} \eta^{-1} (\chi^{P(\eta)} -\eta) \\
			&= -P_{123} \eta^{-1} \sum_{a=1}^h q_{123}^{h-a} \chi^{P(\eta)} + P_{123} \\
			&= -\sum_{a=1}^h q_{123}^{h-a} q_1^{1-A} q_2^{1-B} q_3 ^{1-C} +1
	\end{aligned}
	\eeq 
	There are two subcases:
	\begin{itemize}
		\item case (b1): if $2 \leq A = B= C \leq h$
			(because $h-a+1-A=0 \Rightarrow a=h+1-A \geq 1 \Rightarrow h \geq A$), then we get $-1 +1$, namely no zeros or poles from $A_1$.
		\item case (b2): otherwise, we get exactly one pole from the second term.
	\end{itemize}
\end{itemize}
Since $\Res_{x=\xi} \frac{dx} x \hat a(\xi x^{-1}) = -1$, while $\Res_{x=\xi} \frac{dx}x \hat a(\xi^* x) = +1$,
the desired sign pattern is proven:
terms of type (b1) and (a), which are present exactly when $\xi$ increases $h(\cdot)$ by $0\pmod 2$, do not have poles or zeros in $x^{-1}$ and therefore do not get a minus sign when computing the residue,
while terms of type (b2), which increase $h(\cdot)$ by 1, have a single pole in $x^{-1}$ and therefore get an extra minus sign.

\subsection{The conjecture}
\label{sec:conj}

We conjecture that the instanton partition function can be written as
\beq
\label{conj}
\CalZ_N^{D8} (p,\{q_a\},\{\nu_\alpha\},\{\mu_\alpha\}) =  \operatorname{PE} F \left( q_1, q_2, q_3, q_4, \prod_{\alpha=1}^n \frac {{\nu}_\alpha}{{\mu}_\alpha}, -p \right)
\eeq 
where the plethystic exponent of $f(x_1,\ldots,x_r)$ is
\beq
\operatorname{PE} f(x_1,\ldots,x_r)  = \exp \sum_{m=1}^\infty \frac 1m f(x_1^m,\ldots,x_r^m)
\eeq 
and the argument is
\beq
\label{pletD8}
F  (q_1,q_2,q_3,q_4, s ,p)=
\frac { [q_{12}]  [q_{13}]  [q_{23}]  } { [q_1] [q_2] [q_3] [q_4]}
\frac{ [s] } {[p \sqrt{s}] [p/\sqrt{s}] }
\eeq 
In particular, the dependence on Coulomb branch parameters and masses is only through an overall product factor.
We have checked the conjecture up to the sizes $k=16,7,6,5,4$ for the ranks $N=1,2,3,4,5$ respectively.

The simple dependence on the rank $N$ Coulomb parameters and $N$ masses (or, on the $U(N|N)$ Coulomb parameters), presumably reflects the $U(N|N) \to U(1|1)$ reduction in the non-perturbative setting. 

\subsection{Non-generic parameters and wall-crossing}
\label{sec:wc}

One of the assumptions of the localization computation is that the parameters in $M$ and $N$ are generic, namely they are not related to each other nor to the $q_a$'s.
That is required (roughly speaking) so that the torus action localizes the integral completely,
or equivalently to prove that the tangent space is movable.
If this is not the case, one should first deform the problem such that the parameters are generic enough, and then take the desired degenerate limit.
Our conjecture implies that, when $M$ and $N$ are such that $\prod\limits_{{\alpha} = 1}^{N} {\nu}_{\alpha}  = \prod\limits_{{\alpha} = 1}^{N} {\mu}_{\alpha}  $, then $\CalZ=1$.
This often involves cancellations among many singular terms, and it has been checked in many cases to hold experimentally.
On the other hand, when the parameters are non-generic, one can often deform the action by adding further exact terms,
which localize the path integral to a smaller subset than the generic case.

\subsubsection{Example}

Let us look at an example with rank $2$ and degenerate Coulomb parameters $N= 1 \oplus q_3 q_4$, and masses $M= q_3 \oplus  q_4$.
One can check that indeed the contributions from fixed points cancel for the first few values of $k$;
as $k$ grows, some singular terms appear, but their contributions also cancel, as expected.

On the other hand, since we also have a decomposition $I = I_1 \oplus I_2$ and $\Upsilon= \Upsilon_1 \oplus \Upsilon_2$,
we can add to the gauge fermion $\mathbf \Psi$ the terms
$B_3 I_1 \Upsilon_1^\dagger$ and $B_4 I_1 \Upsilon_2^\dagger$
that are allowed by the particular form of $N$ and $M$, and imply that classical vacua also satisfy the two equations
$B_3 I_1 =0 = B_4 I_1$,
which imply in turn that the first solid partition can only grow along directions 1 and 2.

We observe that in this case the virtual tangent character coincides with that of $U(1)$ $\mathcal N=2^*$ theory in 4d,
and the contribution of the partitions colored by 1 is exactly that of such theory
(they are 2d partitions, growing in the $q_1$, $q_2$ directions).

It would be interesting to see whether it is possible to further modify the action by adding the exact terms
in such a way that partitions colored by $q_3 q_4$
have a vanishing contribution,
in which case we would be left after crossing the wall with the partition function of 4d $U(1)$ $\mathcal N=2^*$ theory.

\subsection{Chern-Simons term}
\label{sec:cs}

We can add to our action a 1d Chern-Simons term
\beq
CS_m = m \int \tr (A_t + \ii \Phi)
\eeq 
for some $m \in \BZ$.
This is gauge invariant and supersymmetric, and upon localization it produces (schematically) a factor of
\beq
e^{CS_m} \to e^{m (u_1 + \cdots u_k)} = (x_1 \cdots x_k)^m \to \left( \prod_{(a,b,c,d) \in \rho} q_1^{a-1} q_2^{b-1} q_3^{c-1} q_4^{d-1} \right)^m \ , 
\eeq 
in other words the measure \cref{res-meas} gets replaced by
\beq
M_{\bz} (\rho) \to M_{\bz} (\rho) \times \left( \prod_{(a,b,c,d) \in \rho} q_1^{a-1} q_2^{b-1} q_3^{c-1} q_4^{d-1} \right)^m
\eeq 
(for simplicity we wrote the $N=1$ case, the higher rank generalization is straightforward).
We could not find a simple plethystic expression in the case $m \neq 0$.

A $D8$-brane acts like a domain wall in type $IIA$ theory, and it separates regions of space where the parameter $m$ differs by one unit.
As a $D0$-brane passes through the $D8$-brane, $m$ jumps by one unit. In addition there is a jump in the slope of the effective inverse string coupling constant.
The $D0$-brane, whose effective mass is proportional to the inverse coupling, feels a force that apparently jumps discontinuously.
However, this discontinuity is balanced by the tension of the fundamental string that is created or destroyed in the process\cite{Bachas:1997kn}.

It would be interesting to connect such CS-terms to some $M,N \to 0,\infty$ limit of our equivariant parameters,
corresponding to moving one or more (anti)$D8$-branes far away at infinity.

\section{Prior conjectures}
\label{sec:d6}

The rank $N=1$ case of \cref{pletD8} is discussed in  Ref.~\cite{Nekrasov:2017cih}. 
Another interesting limit takes us from the theory of multiple $D8$- to the theory of multiple $D6$-branes.

\subsection{\texorpdfstring{$D0-D6$}{D0-D6} brane partition functions}

Consider $k$ $D0$-branes probing the worldvolume of $N$ $D6$-branes in flat space, in the presence of $B$-field.
The Higgs branch coincides with the moduli space $\CalM_{N,k}^\text{D6}$ of representations
of the generalized ADHM quiver associated to the six-dimensional maximally supersymmetric $U(N)$ gauge theory on the $D6$:
it is given by the space of $k \times k$ matrices
$(B^1,B^2,B^3,Y = B^{4})$
and the $k \times N$ matrix $I$
satisfying equations
\beq
\begin{aligned}
	[B^a,B^b] + \varepsilon^{abc} [B_c^\dagger,Y] &= 0\\
	\sum_{a=1}^3 [B^a,B_a^\dagger] + [Y,Y^\dagger] + I \otimes I^\dagger &= r > 0\\
	YI &= 0
\end{aligned}
\eeq 
modulo the action of $U(k)$
\beq
X=(B^1,B^2,B^3,Y,I) \mapsto
g.X = (g B^1 g^{-1}, g B^2 g^{-1}, g B^3 g^{-1}, g Y g^{-1},gI)
\eeq 
The torus action is given by the
$\Omega$ background parameters $q_a = e^{\beta {\ve}_a}$
corresponding to the rotations of the three orthogonal spatial $\BR^2$'s in the worldvolume,
and Coulomb branch parameters ${\nu}_{\alpha} = e^{\ii a_{\alpha}}$ in the Cartan of $U(N)$. Let ${\by} = (\{ q_a\},\{ {\nu}_{\alpha}\} )$. 
We consider the grand canonical partition function $\CalZ_N ^{D6}$ of integrals of some characteristic class $\tau$ in equivariant K-theory
\beq
\CalZ_N ^{D6} (p,\by)
= \sum_{k=0}^\infty p^k \int_{\CalM_{N,k}^\text{D6}} \tau_{N,k}
\eeq 
One can use localization to compute $\CalZ$ as a sum over the fixed points.
Define $u=\operatorname{diag} ({\nu}_1,\ldots,{\nu}_{N})$.
The fixed points are given by matrices for which the torus action
can be undone by a $U(k)$ transformation:
\beq
(q_1 B^1, q_2 B^2, q_3 B^3, (q_1 q_2 q_3)^{-1} Y, I u)
= g(q_1,q_2,q_3,u).X
\eeq 
These are in one-to-one correspondence with the $N$-tuples of plane partitions $\vec{\pi}=(\pi_1,\ldots,\pi_{N})$.
The virtual tangent space is
\beq
\label{tangent6}
T_{\vec \pi} =
\frac {N N^*  - E E^*} {\prod_a \left( 1- q_a^{-1} \right) }
\eeq 
where $N=\sum\limits_{\alpha} {\nu}_{\alpha}$,
$E=N-K \prod\limits_{a=1}^{3} \left( 1-q_a \right)$,
and the character of a plane partition is 
\beq
\operatorname{ch}_\pi (q_1,q_2,q_3)
= \sum_{(a,b,c) \in \pi} q_1^{a-1} q_2^{b-1} q_3^{c-1}
\eeq 
For $N=1$ one can compute $T$ as the difference between deformation and obstruction in perfect obstruction theory.
For $N>1$ one can use a supersymmetric quantum mechanics and apply the same method as \cref{sec:d8}.
Applying the function $\hat a$ to \cref{tangent} gives the result of localization (in Dirac twist) for $\CalZ$.
Concretely, $\CalZ$ is a generating sum over colored plane partitions\cite{Iqbal:2003ds,Cirafici:2008sn}
\beq
\CalZ_N^{D6} (p, {\by} ) = \sum\limits_{k=0}^\infty p^k \sum_{|\vec{\pi}|=k} \mu_{\vec \pi} ( {\by} )
\eeq 
with the weight function determined by $\hat a(T)$
\beq
\mu_{\vec \pi} ( \by )= \prod_{\alpha,\beta=1}^N
\frac
{ \displaystyle \prod_{s \in \pi_\alpha} [q^s \nu_{\alpha\beta}] 
\prod_ {s' \in \pi_\beta} [q^{s-s'} \nu_{\alpha\beta}] \prod_{a=1}^3  [ q_a^{-1} q^{s-s'+1} \nu_{\alpha\beta}] }
{ \displaystyle \prod_{s' \in \pi_\beta} [q^{1-s'} \nu_{\alpha\beta}]
  \prod_{s \in \pi_\alpha} [q^{s-s'+1} \nu_{\alpha\beta}] \prod_{a=1}^3  [q_a q^{s-s'} \nu_{\alpha\beta}] }
\eeq 
We denoted $q^{r+i}=q_1^{r_1+i} q_2^{r_2+i} q_3^{r_3+i}$ for $r \in \pi$, $i \in \BZ$,
and $\nu_{\alpha\beta}=\nu_\alpha \nu_\beta^{-1}$.

To get the $D0-D6$ measure from the $D0-D8$ measure we set
\beq
{\nu}_{\alpha} = q_{4} {\mu}_{\alpha}
\eeq
for all $\alpha = 1, \ldots, N$. Thus, the parameter $s = \prod_{{\alpha}=1}^N \frac{{\nu}_{\alpha}}{{\mu}_{\alpha}}$ in \cref{pletD8} equals $q_{4}^N$ and our conjecture reduces 
to the conjecture of Awata and Kanno\cite{Awata:2009dd} (in particular, it does not depend on the Coulomb moduli $\nu_{\alpha}$, as also observed in Ref.\cite{Awata:2009dd})
and for $N=1$ to the conjecture by one of the authors \cite{Nekrasov:2005bb} proved by Okounkov\cite{Okounkov:2015spn}. In
the rank $N$ case $\CalZ_N ^{D6}$ can be written as a plethystic exponent
(we use $q_4:= (q_1 q_2 q_3)^{-1}$):
\beq
\CalZ_N^{D6} \left( (-1)^N p, {\by} \right)
= \exp \sum_{m=1}^\infty \frac 1m F (q_1^m,q_2^m,q_3^m, q_4^m , q_4^{-Nm},p^m)
\eeq

\paragraph{Remark}

Using \cref{eq:orbchar}  we can write 
\beq
F  (q_1, q_2, q_3, q_4, q_4^{-N}, p) =
\frac 1N \sum_{\ell =0}^{N-1} F (q_1, q_2, q_3,q_4, q_4^{-1}, p^{\frac 1N} e^{2 \pi{\ii} \ell/N})
\eeq
so that $\CalZ_N ^{D6}$ becomes a product of $N$ copies of $\CalZ_{N=1}^{D6}$:
\beq
{\CalZ}_{N}^{D6} ( ( -1)^N  p, q_{1}, q_{2}, q_{3}) = \left( \prod\limits_{{\ell}=0}^{N-1} 
{\CalZ}_{1}^{D6} 
( p^{\frac 1N} e^{\frac{2\pi \ii \ell}{N}}, q_{1}, q_{2}, q_{3}) \right)^{\frac 1N}
\eeq
Using \cref{eq:alechar} we can equally write (use
$z_{1} = q_{123}^{\frac 12} p^{\frac 1{N}}$, $z_{2} = q_{123}^{\frac 12} p^{-\frac 1{N}}$, $z_{1}z_{2} = q_{123}$):
\begin{multline}
F  (q_1, q_2, q_3, q_4, q_4^{-N}, p)  =  \\
-
\sum_{l=1}^{N} \frac{[q_{12}][q_{13}][q_{23}]}{[q_{1}][q_{2}][q_{3}][z_{1}^{N+1-l}z_{2}^{1-l}][z_{1}^{l-N}z_{2}^{l}]} = \\
 \sum_{l=1}^{N}
\frac{[q_{12}][q_{13}][q_{23}]}{[q_{1}][q_{2}][q_{3}][q_{123}^{\frac{N}{2} + 1- l} p ][ q_{123}^{-\frac{N}{2} + l} p^{-1} ]} 
\end{multline}
so that 
\beq
{\CalZ}_{N}^{D6} (p, q_{1}, q_{2}, q_{3}) = \prod_{l=1}^{N}
{\CalZ}_{1}^{D6} \left(  p q_{123}^\frac{N+1 - 2l}2, q_{1}, q_{2}, q_{3} \right)
\eeq
This formula should be compared to the partition function of eleven dimensional supergravity on ${\BR}^{6} \times {\tilde S}_{N}$ fibered over $\BS^{1}$ with the twist on ${\BR}^{6} \approx {\BC}^{3}$ given by
$(q_{1}, q_{2}, q_{3})$ and the compensating rotation of the resolved ALE space ${\tilde S}_{N}$, with $p$ being the hyperK\"ahler $U(1)$ rotation. 
 
In the limit $\ve_{a} \to 0$, keeping ${\ve}_{a}/{\ve}_{b}$ finite, and imposing the Calabi-Yau constraint $\sum\limits_{a=1}^3 {\ve}_a =0$, 
the partition function becomes a product of MacMahon functions $M( {\ldots} )$:
\beq
\CalZ_N^{D6}  \left( (-1)^N p, \{q_a\} \right)  \longrightarrow
\prod_{\ell=0}^{N-1} M \left( p^{\frac 1N} e^{\frac{2 \pi{\ii} \ell}{N}} \right)
^{\frac 1N } = M (p)^{N} \label{eq:cohorb}
\eeq
the last equality being a consequence of the equality of \cref{eq:alechar,eq:orbchar}. 
\Cref{eq:cohorb} agrees with Ref.\cite[eq.~(7.12)]{Cirafici:2012qc} up to a sign redefinition of $p$.

\appendix

\section{Examples}

\subsection{A computation at five instantons}
\label{5inst}

Let us compute the residue contribution for rank $N=1$ from the fixed point
\beq
u_* = ( 0, {\ve}_1,2{\ve}_1,{\ve}_2,{\ve}_1+{\ve}_2)
\eeq 
for which
\beq
Q(u_*) = \{ e_1, e_{21}, e_{31}, e_{42}, e_{52}, e_{53} \}
\eeq 
Let us take a small $\delta >0$ and
\beq
\xi = (1+5\delta) e_1+ (1+4\delta) e_2+ (1+3\delta) e_3+ (1+2\delta) e_4+ (1+\delta) e_5
\eeq 
By computing products $\xi \cdot e_i$ we see that the only flag contributing to the residue is
\beq
F = \langle e_1 \rangle,  \langle e_1,e_{21} \rangle,  \langle e_1,e_{21},e_{31} \rangle,  \langle e_{42},e_{31},e_{21},e_1 \rangle
\eeq 
which has $\nu(F)=+1$. The iterated residue at $F$ of $\chi_5$ is equal to its iterated residue at the flag given by our choice of ordering, both being
\beq
\begin{aligned}
& \Res_F \chi_5 = -\frac{(-1 + {\mu}) ({\mu} - q_1) ({\mu} - q_1^2) ({\mu} - q_2) ({\mu} - q_1 q_2) (-1 + q_1 q_2)^2  }
{ (-1 + q_1)^3 (1 + q_1) (q_1 - q_2)^2 (q_1^2 - q_2) (q_1^3 - q_2) (-1 + q_2)^2} \\
& \quad \times \frac{(1 + q_1 q_2) (-1 + q_1^2 q_2) (-1 + q_1^3 q_2) (-1 + q_1 q_2^2)  (-1 + q_1 q_3)^2 (-1 + q_1^2 q_3) }
{  (q_1 - q_2^2) (q_1 - q_3) (q_1^2 - q_3) (q_2 - q_3) (q_1 q_2 - q_3) (-1 + q_3) (q_1 + q_2)} \\
& \quad \times \frac{ (-q_2 + q_1^2 q_3) (-q_2 + q_1^3 q_3)  (q_1 - q_2 q_3) (-1 + q_2 q_3)^2 (q_1 - q_2^2 q_3) (q_1^2 -  q_2^2 q_3)}
{(-1 + q_1 q_2 q_3) (-1 + q_1^2 q_2 q_3) (-1 + q_1^3 q_2 q_3) (-1 + q_1 q_2^2 q_3) (-1 + q_1^2 q_2^2 q_3)}
\end{aligned}
\eeq

\subsection{A check of the conjecture at two instantons}

\label{example}

Let us check the simplest non-trivial case,  $k=2$ instantons at rank $N=2$.
There are four solid partitions of size 2, with character $1+q_i$ for $i =1, \ldots, 4$, and one solid partition of size 1, with character 1.
Therefore the pairs contributing to the two instanton sector are $(1+q_i, 0)$, $(0,1+q_i)$ and $(1,1)$. 
The partition function is
\beq
\label{2inst}
\CalZ_{k=2,N=2}^{D8} = z_{(1,1)} + \sum_{i=1}^4 ( z_{(1+q_i, 0)} +z_{(0,1+q_i) } )
\eeq 
with
\beq
\begin{aligned}
z_{(1,1)} &=  - \frac	
{({\mu}_1 - {\nu}_1) ({\mu}_2 - {\nu}_1) ({\mu}_1 - {\nu}_2) ({\mu}_2 - {\nu}_2) (-1 + q_1 q_2)^2}
{{\mu}_1 {\mu}_2 (-1 + q_1)^2 (-{\nu}_2 + {\nu}_1 q_1) (-{\nu}_1 + {\nu}_2 q_1) (-1 + q_2)^2} \\
& \times \frac { (-{\nu}_2 + {\nu}_1 q_1 q_2) ({\nu}_1 - {\nu}_2 q_1 q_2)(-1 + q_1 q_3)^2 (-{\nu}_2 + {\nu}_1 q_1 q_3)}
{ ({\nu}_2 - {\nu}_1 q_2) (-{\nu}_1 + {\nu}_2 q_2)(-1 + q_3)^2 ({\nu}_2 - {\nu}_1 q_3) (-{\nu}_1 + {\nu}_2 q_3) } \\
& \times \frac  {({\nu}_1 - {\nu}_2 q_1 q_3) (-1 + q_2 q_3)^2 (-{\nu}_2 + {\nu}_1 q_2 q_3) ({\nu}_1 - {\nu}_2 q_2 q_3)}
{(-1 + q_1 q_2 q_3)^2 (-{\nu}_2 + {\nu}_1 q_1 q_2 q_3) (-{\nu}_1 + {\nu}_2 q_1 q_2 q_3)}
\end{aligned}
\eeq 
\beq
\begin{aligned}
z_{(1+q_1,0)} &= - \frac	
{({\mu}_1 - {\nu}_1) ({\mu}_2 - {\nu}_1) {\nu}_2 ({\mu}_1 - {\nu}_1 q_1) ({\mu}_2 - {\nu}_1 q_1) (-1 + q_1 q_2)}
{{\mu}_1 {\mu}_2 {\nu}_1 ({\nu}_1 - {\nu}_2) (-1 + q_1)^2 (1 + q_1) (-{\nu}_2 + {\nu}_1 q_1)(q_1 - q_2)} \\
& \times \frac {(-1 +q_1^2 q_2) (-1 + q_1 q_3) (-1 + q_1^2 q_3) (q_1 - q_2 q_3) (-1 + q_2 q_3)}
{(-1 +q_2) (q_1 - q_3) (-1 + q_3) (-1 + q_1 q_2 q_3) (-1 + q_1^2 q_2 q_3)}
\end{aligned}
\eeq 
\beq
\begin{aligned}
z_{(1+q_2,0)} &= - \frac	
{({\mu}_1 - {\nu}_1) ({\mu}_2 - {\nu}_1) {\nu}_2 ({\mu}_1 - {\nu}_1 q_2) ({\mu}_2 - {\nu}_1 q_2) (-1 + q_1 q_2)}
{{\mu}_1 {\mu}_2 {\nu}_1 ({\nu}_1 - {\nu}_2) (-1 + q_1) (q_1 - q_2) (-1 + q_2)^2 (1 + q_2)} \\
& \times \frac {(-1 +q_1 q_2^2) (-1 + q_1 q_3) (-q_2 + q_1 q_3) (-1 + q_2 q_3) (-1 + q_2^2 q_3)}
{(-{\nu}_2 +  {\nu}_1 q_2) (q_2 - q_3) (-1 + q_3) (-1 + q_1 q_2 q_3) (-1 + q_1 q_2^2 q_3)}
\end{aligned}
\eeq 
\beq
\begin{aligned}
z_{(1+q_3,0)} &= \frac
{({\mu}_1 - {\nu}_1) ({\mu}_2 - {\nu}_1) {\nu}_2 (-1 + q_1 q_2) (q_1 q_2 - q_3) ({\mu}_1 - {\nu}_1 q_3)}
{{\mu}_1 {\mu}_2 {\nu}_1 ({\nu}_1 - {\nu}_2) (-1 + q_1) (-1 + q_2) (q_1 - q_3) (q_2 -  q_3)} \\
& \times \frac {({\mu}_2 - {\nu}_1 q_3) (-1 + q_1 q_3) (-1 + q_2 q_3) (-1 + q_1 q_3^2) (-1 + q_2 q_3^2)}
{(-1 + q_3)^2 (1 + q_3) (-{\nu}_2 + {\nu}_1 q_3) (-1 + q_1 q_2 q_3) (-1 + q_1 q_2 q_3^2)}
\end{aligned}
\eeq 
\beq
\begin{aligned}
z_{(1+q_4,0)} &= \frac	
{({\mu}_1 - {\nu}_1) ({\mu}_2 - {\nu}_1) {\nu}_2 (-1 + q_1 q_2) (-1 + q_1 q_3) (-1 + q_2 q_3)}
{{\mu}_1 {\mu}_2 {\nu}_1 ({\nu}_1 - {\nu}_2) (-1 + q_1) (-1 + q_2) (-1 + q_3) (-1 + q_1 q_2 q_3)^2}\\
& \times \frac { (-{\nu}_1 + {\mu}_1 q_1 q_2 q_3)(-{\nu}_1 + {\mu}_2 q_1 q_2 q_3) (-1 + q_1^2 q_2^2 q_3) (-1 + q_1^2 q_2 q_3^2)}
{ (1 + q_1 q_2 q_3)({\nu}_1 - {\nu}_2 q_1 q_2 q_3) (-1 + q_1^2 q_2 q_3) (-1 + q_1 q_2^2 q_3)}\\
& \times \frac
{ -1 + q_1 q_2^2 q_3^2 }
{ -1 + q_1 q_2 q_3^2 }
\end{aligned}
\eeq 
\beq
\begin{aligned}
z_{(0,1+q_1)} &= -\frac	
{{\nu}_1 ({\mu}_1 - {\nu}_2) (m-2 - {\nu}_2) ({\mu}_1 - {\nu}_2 q_1) ({\mu}_2 - {\nu}_2 q_1) (-1 + q_1 q_2)}
{{\mu}_1 {\mu}_2 ({\nu}_1 - {\nu}_2) {\nu}_2 (-1 + q_1)^2 (1 + q_1) ({\nu}_1 - {\nu}_2 q_1) (q_1 - q_2)} \\
& \times \frac {(-1 +q_1^2 q_2) (-1 + q_1 q_3) (-1 + q_1^2 q_3) (q_1 - q_2 q_3) (-1 + q_2 q_3)}
{(-1 + q_2) (q_1 - q_3) (-1 + q_3) (-1 + q_1 q_2 q_3) (-1 + q_1^2 q_2 q_3)}
\end{aligned}
\eeq 
\beq
\begin{aligned}
z_{(0,1+q_2)} &= - \frac
{{\nu}_1 ({\mu}_1 - {\nu}_2) ({\mu}_2 - {\nu}_2) ({\mu}_1 - {\nu}_2 q_2) ({\mu}_2 - {\nu}_2 q_2) (-1 + q_1 q_2)}
{{\mu}_1 {\mu}_2 ({\nu}_1 - {\nu}_2) {\nu}_2 (-1 + q_1) (q_1 - q_2) (-1 + q_2)^2 (1 + q_2)} \\
& \times \frac
{(-1 +q_1 q_2^2) (-1 + q_1 q_3) (-q_2 + q_1 q_3) (-1 + q_2 q_3) (-1 + q_2^2 q_3)}
{({\nu}_1 - {\nu}_2 q_2) (q_2 - q_3) (-1 + q_3) (-1 + q_1 q_2 q_3) (-1 + q_1 q_2^2 q_3)}
\end{aligned}
\eeq 
\beq
\begin{aligned}
z_{(0,1+q_3)} &= \frac	
{{\nu}_1 ({\mu}_1 - {\nu}_2) ({\mu}_2 - {\nu}_2) (-1 + q_1 q_2) (q_1 q_2 - q_3) ({\mu}_1 - {\nu}_2 q_3)}
{ {\mu}_1 {\mu}_2 ({\nu}_1 - {\nu}_2) {\nu}_2 (-1 + q_1) (-1 + q_2) (q_1 - q_3) (q_2 - q_3)} \\
& \times \frac
{({\mu}_2 - {\nu}_2 q_3) (-1 + q_1 q_3) (-1 + q_2 q_3) (-1 + q_1 q_3^2) (-1 + q_2 q_3^2)}
{(-1 + q_3)^2 (1 + q_3) ({\nu}_1 - {\nu}_2 q_3) (-1 + q_1 q_2 q_3) (-1 + q_1 q_2 q_3^2)}
\end{aligned}
\eeq 
\beq
\begin{aligned}
z_{(0,1+q_4)} &= \frac
{{\nu}_1 ({\mu}_1 - {\nu}_2) ({\mu}_2 - {\nu}_2) (-1 + q_1 q_2) (-1 + q_1 q_3) (-1 + q_2 q_3) }
{{\mu}_1 {\mu}_2 ({\nu}_1 - {\nu}_2) {\nu}_2 (-1 + q_1) (-1 + q_2) (-1 + q_3) (-1 + q_1 q_2 q_3)^2} \\
& \times \frac
{(-{\nu}_2 + {\mu}_1 q_1 q_2 q_3)(-{\nu}_2 + {\mu}_2 q_1 q_2 q_3) (-1 + q_1^2 q_2^2 q_3) (-1 + q_1^2 q_2 q_3^2)}
{(1 + q_1 q_2 q_3) (-{\nu}_2 + {\nu}_1 q_1 q_2 q_3) (-1 + q_1^2 q_2 q_3) (-1 + q_1 q_2^2 q_3) }\\
& \times \frac
{ -1 + q_1 q_2^2 q_3^2 }
{-1 + q_1 q_2 q_3^2}
\end{aligned}
\eeq 
One can check that \cref{2inst} has the plethystic representation \cref{conj}.

\section{Definitions}
\label{defs}

We call a collection $\mathcal A$ of elements $\mathcal Q_i \in \BR^{k*}$ a hyperplane arrangement.
We call $\mathcal A$ projective if it lies in a half space of $\BR^{k*}$.
If $\mathcal A$ is projective, let $\Cone_\text{sing}(\mathcal A)$ be the union of cones generated by all subsets of $\mathcal A$ with $k-1$ elements.
Then each connected component of $\Cone (\mathcal A) \setminus \Cone_\text{sing}(\mathcal A)$ is called a chamber.

Denote $\mathcal {FL}(\mathcal A)$ the finite set of flags
\beq
F = [ F_0 =\{0\}\subset F_1 \subset \ldots \subset F_k = \BR^{k*}]
\eeq 
where $\dim F_j=j$ and $\mathcal A \cap F_j$ spans $F_j$ for every $j$.
For $F \in \mathcal {FL}(\mathcal A)$, define
\beq
\kappa_j^F = \sum_{\mathcal Q_i \in F_j} \mathcal Q_i, \qquad j=1,\ldots, k 
\eeq 
\beq
s^+ (F,\mathcal A) = \left\{ \sum_{j=1}^k m_j \kappa_j^F | m_j \in \BR_{\geq 0} \right\}
\eeq 
If the $k$ vectors $\kappa^F_j$ are linearly independent, define $\nu (F)=\pm 1$ depending on whether the sequence $[\kappa^F_1,\ldots, \kappa^F_k]$ is positively oriented.

For $F \in \mathcal {FL}(\mathcal A)$, take a positively oriented basis $\{ \gamma_i^F \}_{i=1}^k$ such that $\{ \gamma_i^F \}_{i=1}^j$ generates $F_j$;
then the iterated residue at $F$ is defined using such a basis.

For $\xi \in \BR^{k*}$, denote
\beq
\mathcal {FL}^+ (\mathcal A, \xi) = \{ F \in \mathcal {FL}(\mathcal A) | \xi \in s^+ (F,\mathcal A) \}
\eeq 
We say that $\xi$ is $\mathcal{FL}(\mathcal A)^+$-regular if for every $F\in \mathcal {FL}^+ (\mathcal A, \xi)$,
in any decomposition $\xi = \sum_{j=1}^k m_j \kappa^F_j$ all the coefficients $m_j$ are non-zero.

\section{ALE computations}
\label{sec:ale}

The character of the space of sections of the spin bundles over ${\BC}^{2}$ is equal to the equivariant index of Dirac operator, which is equal to
\beq
{\chi}_{{\BC}^{2}}(z_{1}, z_{2}) =  \frac{1}{[z_{1}][z_{2}]} \equiv \frac{\sqrt{z_{1}z_{2}}}{(1-z_{1})(1-z_{2})}
\eeq
where $z_{1}, z_{2}$ are the weights of the two holomorphic coordinates.
Now let us perform an orbifold by ${\BZ}_{N} \subset SU(2)$, which acts via $(z_{1}, z_{2}) \mapsto ({\varpi} z_{1}, {\varpi}^{-1} z_{2})$, ${\varpi} = {\exp} \, \frac{2\pi\ii}{N}$:
\beq
{\chi}_{{\BC}^{2}/{\BZ}_{N}} =  \frac{1}{N} \sum_{{\ell}=0}^{N-1} \frac{\sqrt{z_{1}z_{2}}}{(1-{\varpi}^{\ell} z_{1})(1-{\varpi}^{-{\ell}} z_{2})}
\eeq
Using the simple identity
\beq
\frac 1N \sum_{\ell=0}^{N-1} \frac 1 {1- \gamma {\varpi}^{\ell}} = \frac 1 {1 -\gamma^N}\ ,
\eeq 
for any complex $\gamma$, which is not an $N$-th root of unity, we compute:
\beq
{\chi}_{{\BC}^{2}/{\BZ}_{N}}(z_{1}, z_{2})  = \frac{[z_{1}^{N}z_{2}^{N}]}{[z_{1}z_{2}][z_{1}^{N}][z_{2}^{N}]} = {\chi}_{{\BC}^{2}} \left( z_{1}^{N}, z_{2}^{N}\right) \sum\limits_{m=-\frac{N-1}{2}}^{\frac{N-1}{2}} (z_{1}z_{2})^{m}  
\label{eq:orbchar}
\eeq
We call \cref{eq:orbchar} an orbifold version of the character. 
One the other hand, the orbifold ${\BC}^{2}/{\BZ}_{N}$ admits a toric resolution ${\tilde S}_{N}$, with the two-dimensional torus action, which asymptotically is the action of the torus on ${\BC}^2$ projected down to the quotient by ${\BZ}_{N}$. The resolved space ${\tilde S}_{N}$ has $N$ isolated fixed points $p_{l}$, $l = 1, \ldots , N$ with local weights given by:
\beq
{\chi}_{T_{p_{l}}} = z_{1}^{N+1-l} z_{2}^{1-l}  + z_{1}^{l-N} z_{2}^{l} 
\eeq
The index of Dirac operator on ${\tilde S}_{N}$ is computed by the fixed point formula, giving:
\beq
{\chi}_{{\BC}^{2}/{\BZ}_{N}}(z_{1}, z_{2}) = \sum_{l=1}^{N} {\chi}_{{\BC}^{2}}\left( z_{1}^{N+1-l} z_{2}^{1-l}, z_{1}^{l-N} z_{2}^{l} \right)
\label{eq:alechar}
\eeq
We call \cref{eq:alechar} a blown up version of the character. The equality of \cref{eq:alechar,eq:orbchar} can be checked directly. However it is amusing to observe that the equality is a consequence of the compactness of the fibers of the projection
${\tilde S}_{N} \longrightarrow {\BC}^{2}/{\BZ}_{N}$ (see Ref.\cite{Nekrasov:2004vw} for a similar discussion). 

Note that for each $l$ the product of the two weights is equal to $z_{1}z_{2}$. This is a reflection of the Calabi-Yau nature of the resolution ${\tilde S}_{N}$.

\bibliographystyle{JHEP}
\bibliography{biblio-inspire}

\end{document}